\documentstyle[12pt]{article}
\begin{document}
\begin{center}
{\large \bf
Spin Correlations in Top Quark Production and the Top Quark Mass
\\ }
\vspace{5mm}
Gary R. Goldstein
\\
\vspace{3mm}
{\small\it
Department of Physics, Tufts University, Medford, MA 02155,  USA
\\ }
\end{center}

\begin{center}
ABSTRACT

\vspace{5mm}
\begin{minipage}{130 mm}
\small
Top-antitop quark pairs produced at the Tevatron have a sizeable spin
correlation. That correlation feeds into the angular distribution of the
decay products, particularly in the dilepton channel. Including the
expected correlation in an overall analysis of a handful of actual
dilepton events continues to favor a lower top mass (centered on 155 GeV)
than the single lepton events. 
\end{minipage}
\end{center}

In 1992 a
phenomenological study of the top quark's production and decay
characteristics was presented[1], based on the observation that the
contemporaneous lower limit on the mass allowed the top quark to decay
into an on-shell W boson and bottom quark. The kinematics of this two-body 
decay constrain the subsequent final state kinematics, whether the W 
decays leptonically or into a quark-antiquark pair.
In a $\bar{p}p$ collision the principal mechanism for producing a heavy
pair of quarks is either through the light quark pair annihilation 
or gluon fusion. 
The decay chain $t \rightarrow W^+ + b$ 
and $W^+ \rightarrow l^+ + \nu_l$ for the top and the charge conjugate 
chain for the antitop, leads to two energetic leptons and two missing 
neutrinos in the final state - the dilepton mode.
For a given top mass the observed configuration can
be assigned a probability, since we know {\it a priori\/} what the
probability is for that configuration to be realized.
Hence, a likelihood can be assigned to
each event, as a function of the top mass, for that event to correspond to
a {\it bona fide\/} top event. 

This method for determining the likelihood and most probable mass for a
hypothetical top production via the dilepton or single, unilepton mode
[1,2] (see also [3]) was 
applied to the meager published 
%%%%%%%%%%%%%%%%%%%template stuff follows
%
%
% the following 8 cm wide figure will be placed in 
% a 8 cm wide box on the right side of the page 
%
%\begin{wrapfigure}{r}{8cm}
%\epsfig{figure=goldstein1.eps,width=8cm}
%{\small Figure 1: The Log of the joint probabilities vs. $m_t$ given in
%Ref. [4].}
%\end{wrapfigure}
%
%%%%%%%%%%%%%%%%%%%%%%%%%%%%%
data from CDF. For simplicity, the
possible spin correlation of the top pair was ignored in defining the
likelihood. In Fig.1 the results for that analysis[4]  
are shown. The joint
probability was computed for the dileptons and unileptons separately. The
results of the analysis confirmed the CDF results for their seven
published single lepton events and favor a mass near 175
GeV[5]. However, the three dilepton events, whose measured momenta
were available, were analysed also (and the one D0 event, that 
the group had analysed[6], was included in the joint probability) and  
led to a lower mass of $156\pm 8$ GeV. This lower value is consistent with  
results (using a modified Dalitz and Goldstein method) presented subsequently 
by the D0 group for 5 dilepton events[7]. 

Is this result significant? Because of lower backgrounds and fewer jets, 
the dilepton events should be more reliable indicators of top production.
Can the mass determination for the dileptons be sharpened? This can be 
accomplished by including the spin correlations 
between the top and antitop in the likelihood function. 
A preliminary study was begun several years ago (with R.H. Dalitz and      
K. Sliwa) At the tree level of QCD 
for light quark-antiquark annihilation into heavy quark
pairs, the spins of the heavy quarks tend to be aligned, while for
gluon fusion the heavy quark spins tend toward anti-alignment. It is known, 
however, that the gluon fusion mechanism is supressed relative to 
light quark annihilation for the Tevatron energy and the expected range of 
top masses[8]. Hence, the alignment of spins is preferred.

The top spin correlations can be expressed as a double density matrix
in a direct product form[9].
The resulting
spin correlations are transmitted to the decay products. Ignoring
the b-jets (whose polar angle relative to the lepton direction is fixed by 
kinematics), the correlations between the lepton directions and the parent
top spin (in the top rest frame) produce correlations between the lepton
directions, expressed here as a weighting factor:
\begin{eqnarray}
W(\theta,p,p_{\bar l},p_l) & = & \frac{1}{4}
\left\{1+[sin^2\theta([p^2+m^2](\hat{p}_{\bar l})_x(\hat{p}_l)_{\bar{x}} +
[p^2-m^2] (\hat{p}_{\bar l})_y(\hat{p}_l)_{\bar{y}}\right. \nonumber\\
                           &   & \mbox{} - 2 m p cos\theta
sin\theta((\hat{p}_{\bar l})_x(\hat{p}_l)_{\bar{z}} + (\hat{p}_{\bar
l})_z(\hat{p}_l)_{\bar{x}})  + ([p^2-m^2]\nonumber\\
                           &   & \mbox{} \left. +[p^2+m^2] cos^2\theta)
(\hat{p}_{\bar l})_z(\hat{p}_l)_{\bar{z}}] 
            / [(p^2+m^2)+(p^2-m^2)cos^2\theta]\right\} 
\nonumber
\end{eqnarray} 
where $\theta$ is the top quark production
angle in the quark-antiquark CM frame, $p$ is the quark CM momentum,
$\hat{p}_{\bar{l}}$ is the $l^+$ momentum direction in the top rest frame
and $\hat{p}_l$ is the corresponding $l^-$ direction in the antitop rest
frame. For this $q \bar{q}$ case the leptons will tend to be oppositely
oriented in their respective top rest frames. When boosted back to the
subprocess center-of-mass frame the same conclusion will hold, namely that
$q \bar{q}$ annihilation favors a large opening angle between the leptons.
Finally, after the subprocess is folded back into the 
$p \bar{p}$ annihilation with
the proton structure functions, the effect gets diluted, but not
completely. 

To study the importance of the spin correlation, a Monte Carlo generator 
of dilepton events with spin correlated
%%%%%%%%%%%%%%%%%%%template stuff follows
%
%
% the following 8 cm wide figure will be placed in 
% a 8 cm wide box on the right side of the page 
%
%\begin{wrapfigure}{r}{8cm}
%\epsfig{figure=goldstein2.eps,width=8cm}
%{\small Figure 2: Probability distribution for Monte Carlo events with and 
%without spin correlations.}
%\end{wrapfigure}
%
%%%%%%%%%%%%%%%%%%%
top quark pairs was produced. The analysis procedure is
alterred to include the spin correlation weight in the
determination of the overall Bayesian probability. When this is done for
all the events, the resulting mass distribution is more sharply peaked
than it is for the uncorrelated events as Fig.2 corroborates.

The new likelihood function, including the spin correlation weighting 
factor, is applied to the three published CDF events. The resulting 
probability distribution shifts the peaks from the previous analysis, but 
%%%%%%%%%%%%%%%%%%%template stuff follows
%
%
% the following 8 cm wide figure will be placed in 
% a 8 cm wide box on the right side of the page 
%
%\begin{wrapfigure}{r}{8cm}
%\epsfig{figure=goldstein3.eps,width=8cm}
%{\small Figure 3: One CDF dilepton event[5] probability distribution with spin 
%correlations included. See Ref. [4] for uncorrelated case.}
%\end{wrapfigure}
%
%%%%%%%%%%%%%%%%%%%%%%%%%%%%%
does not alter the conclusion that the \underline{dilepton events indicate a 
lower mass for the top quark}. Fig.3 shows one of the events, illustrating 
that there is some gain in the sharpness of the distribution. 

Without a larger 
sample of events it is not possible to draw any further conclusions, but 
the results here should encourage the application of the spin correlation  
weighting to the analyses of the existing dilepton data.

%\vspace{0.2cm}
%\vfill
%%%%%%%How to prevent double spacing after each item?????????
{\small\begin{description}
\item{[1]}   R. H. Dalitz and Gary R. Goldstein, Phys. Lett.
  {\bf B287}, 225 (1992); Phys. Rev. {\bf D45}, 1531 (1992);
   Int. J. Mod. Phys. {\bf A9}, 635 (1994).
\item{[2]} Gary R. Goldstein, K. Sliwa and R. H. Dalitz, Phys. Rev.
  {\bf D47}, 967 (1993).
\item{[3]}   K. Kondo, {\it et al.} J. Phys. Soc. Japan, {\bf 57}, 4126 (1988);
  {\it ibid}, {\bf 60}, 836 (1991); {\it ibid}, {\bf 62}, 1177 (1993).
\item{[4]} R. H. Dalitz and G. R. Goldstein, 
     ``{\it Top mass analysis for the reported top-antitiop production and 
     decay events}'', Tufts preprint TUFTS TH-95-G01 (1995).
\item{[5]}   F. Abe {\it et al.} (CDF Collaboration), Phys. Rev. Lett.
  {\bf 73}, 225 (1994); Phys. Rev. {\bf D50}, 2966 (1994).
\item{[6]}  S. Abachi {\it et al.} (D0 Collaboration), Phys. Rev. Lett.
  {\bf 74}, 2632 (1995).
\item{[7]} E. Varnes (for D0 Collaboration), {\it in:}
  Proceedings of the Division of Particles and Fields Meeting 1996
     Minneapolis, Minn. (Aug.1996), to be published.
\item{[8]} D. Chang, S.-C. Lee, and A. Soumarokov, Phys. Rev. Lett. 
     {\bf 77}, 1218 (1996).
\item{[9]} R. H. Dalitz, G. R. Goldstein, R. Marshall, Phys. Lett. 
   {\bf B215}, 783 (1988); and
    K. Chen, G. R. Goldstein, R. L. Jaffe, X. Ji, Nucl. Phys. 
    {\bf B445}, 380 (1995).
\end{description}}

{\small{\bf Figure Captions}\newline
Figure 1: The Log of the joint probabilities vs. $m_t$ given in
Ref. [4].\newline
Figure 2: Averaged probability distribution for Monte Carlo events with and
without spin correlations.\newline
Figure 3: One CDF dilepton event[5] probability distribution with spin
correlations included. See Ref. [4] for uncorrelated case.}

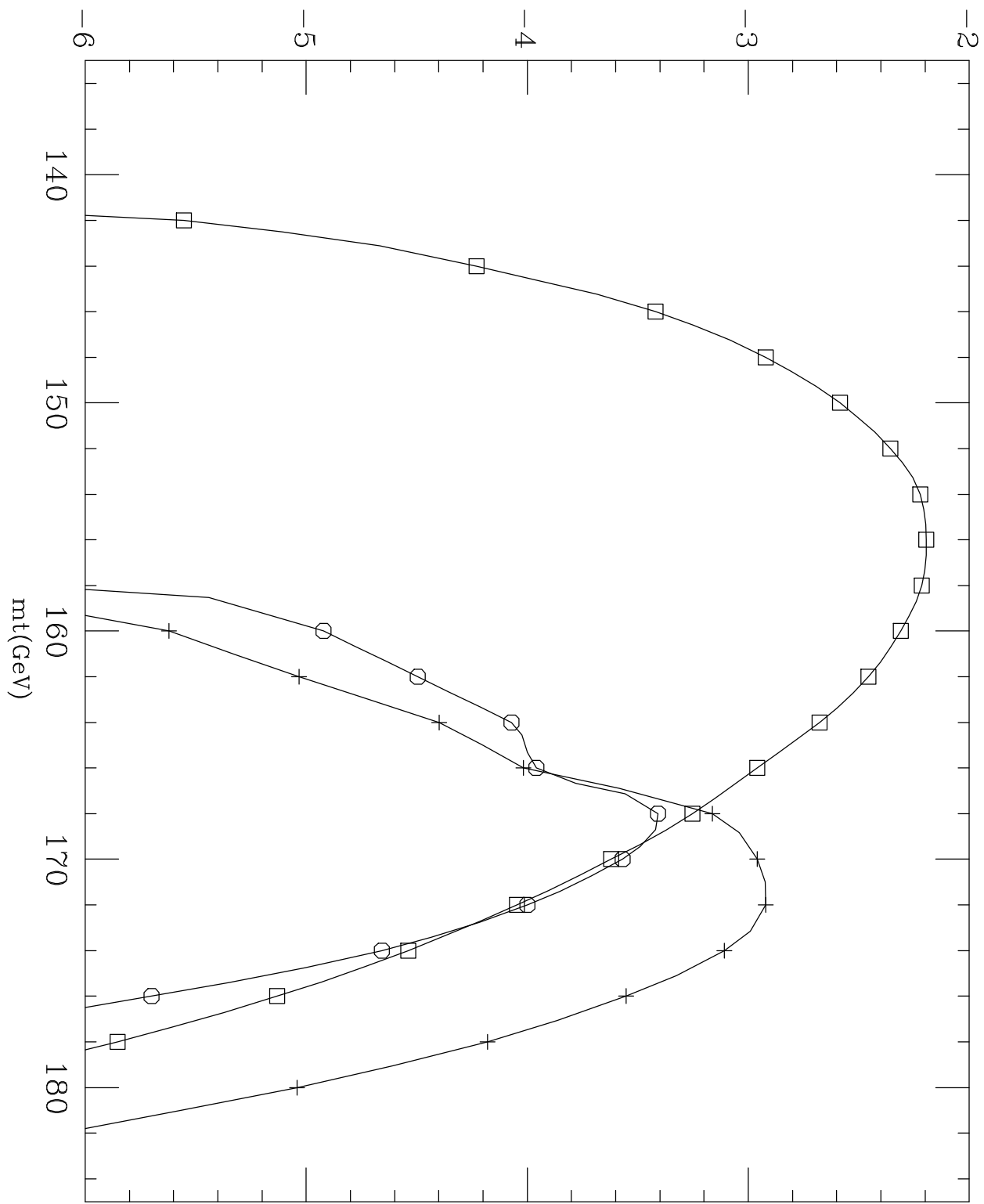
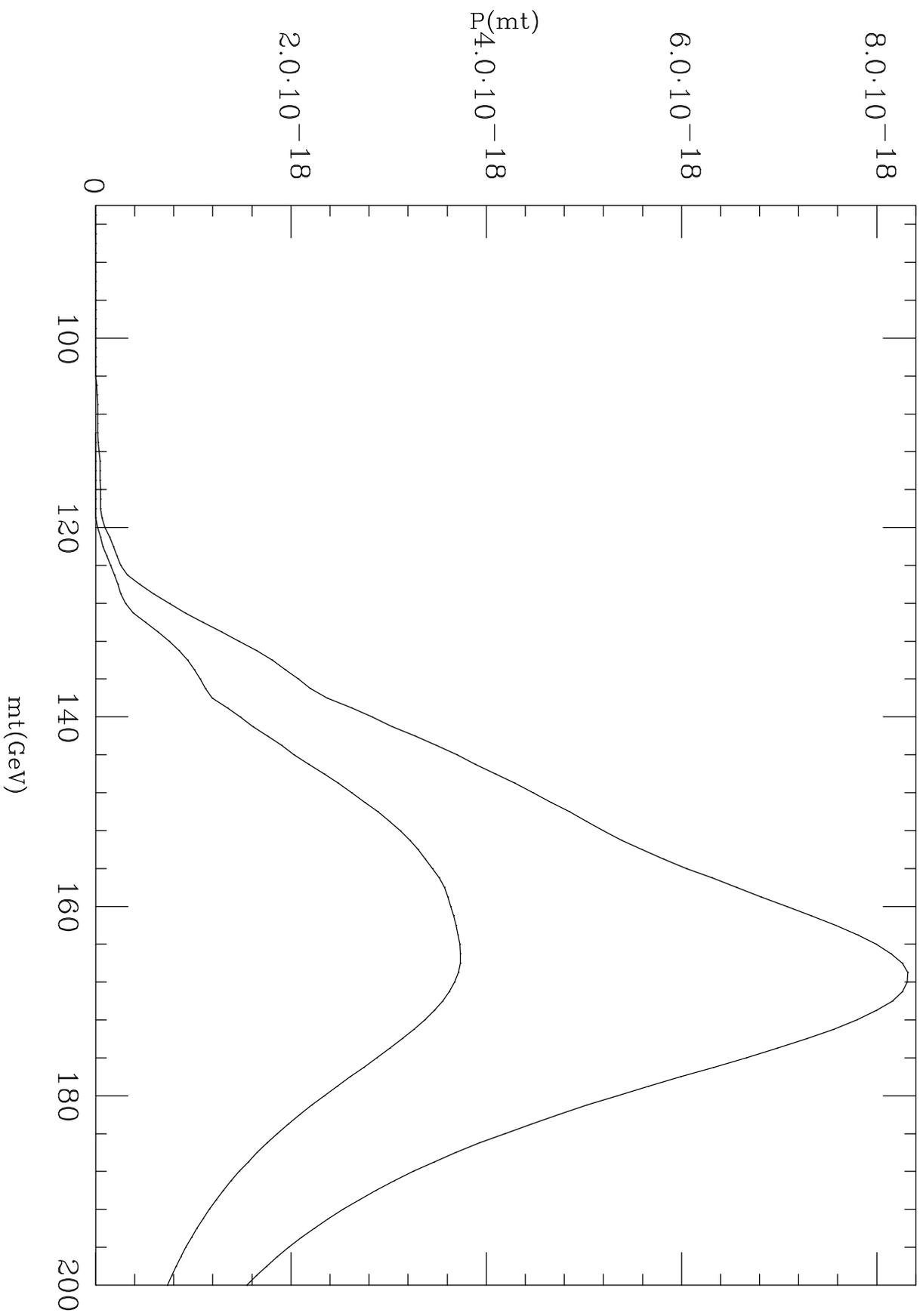
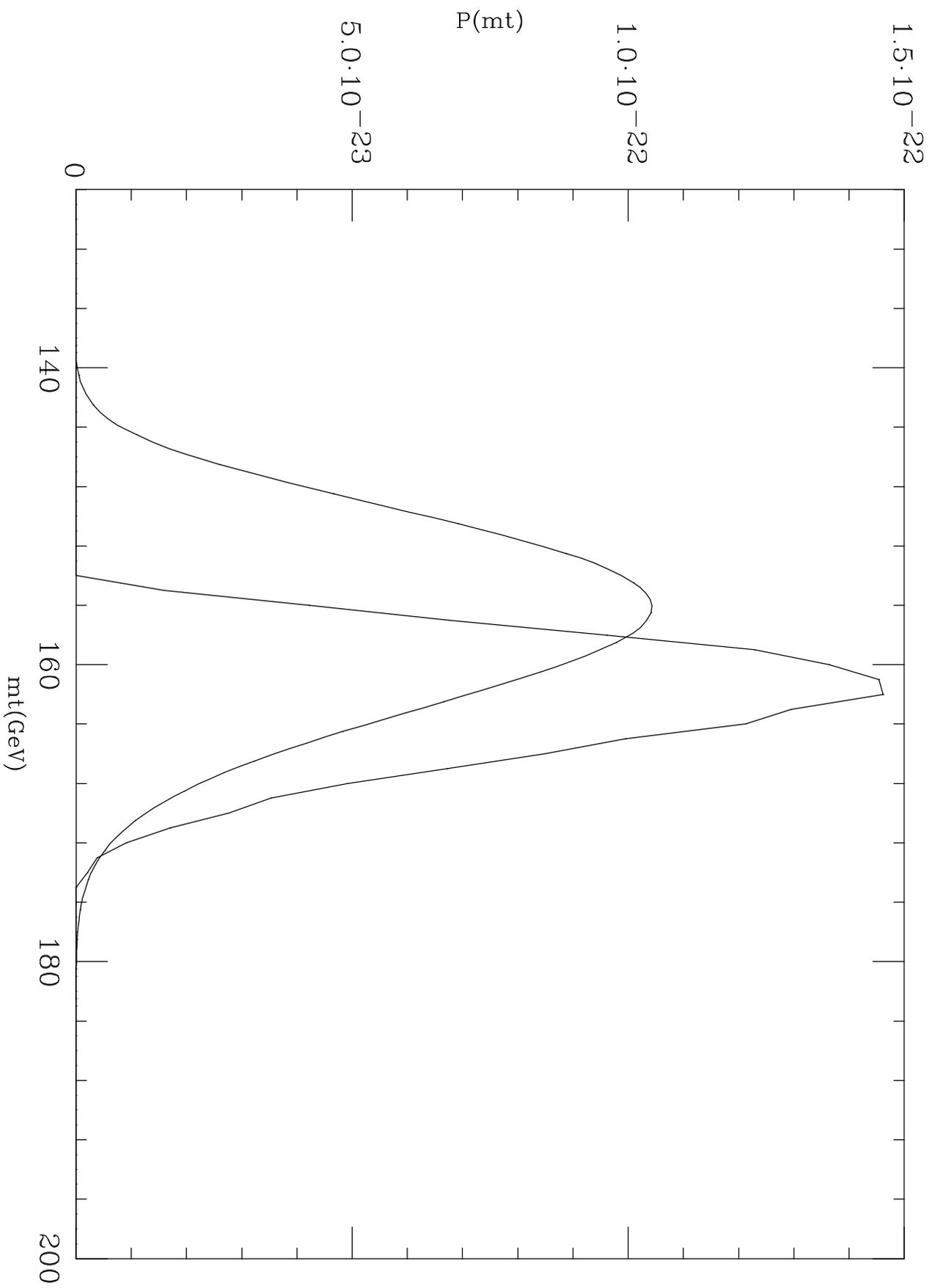
\end{document}